\documentclass[namedreferences]{solarphysics}

\usepackage[hyperref,optionalrh,showbiblabels]{spr-sola-addons} 
\usepackage{graphicx}        
\usepackage{color}           
\usepackage{breakurl}        

\chardef\us=`\_

\begin{document}

\begin{article}
\begin{opening}

\title{Four decades of advances from MSDP to S4I and SLED imaging spectrometers}

\author{P.~\surname{Mein}$^{1}$
\sep
        J.-M.~\surname{Malherbe}$^{2}$
\sep
        F.~\surname{Say\`{e}de}$^{3}$
\sep
        P.~\surname{Rudawy}$^{4}$
\sep
        K.J.H.~\surname{Phillips}$^{5}$
\sep
        F.P.~\surname{Keenan}$^{6}$
        }

\runningauthor{P. Mein, J.M. Malherbe, F. Say\`{e}de, P. Rudawy, K.
Phillips, F. Keenan }

\runningtitle{Advances from MSDP to S4I and SLED imaging
spectrometers}

\institute{ $^{1}$ LESIA, Observatoire de Paris, 92195 Meudon, and
PSL Research University, France,
                 email: \url{Pierre.Mein@club-internet.fr}\\
$^{2}$ LESIA, Observatoire de Paris, 92195 Meudon, and PSL Research
University, France,
                 email: \url{Jean-Marie.Malherbe@obspm.fr}, ORCID: 0000-0002-4180-3729\\
                 $^{3}$ GEPI, Observatoire de Paris, 92195 Meudon, and PSL Research
University, France,
                 email: \url{Frederic.Sayede@obspm.fr}\\
              $^{4}$ Wroc{\l}aw University, Poland,
               email: \url{rudawy@astro.uni.wroc.pl}\\
              $^{5}$ Earth Sciences Department, Natural History Museum, London SW75BD, United Kingdom,
               email: \url{kennethjhphillips@yahoo.com}, ORCID: 0000-0002-3790-990X\\
              $^{6}$ Astrophysics Research Centre, School of Mathematics and Physics, Queen's University Belfast, United Kingdom,
               email: \url{f.keenan@qub.ac.uk}
               }

\begin{abstract}
The Multichannel Subtractive Double Pass (MSDP) is an imaging
spectroscopy technique, which allows observations of spectral line
profiles over a 2D field of view with high spatial and temporal
resolution. It has been intensively used since 1977 on various
spectrographs (Meudon, Pic du Midi, the German Vacuum Tower
Telescope, THEMIS, Wroc{\l}aw). We summarize previous developments
and describe the capabilities of a new design that has been
developed at Meudon and that has higher spectral resolution and
increased channel number: Spectral Sampling with Slicer for Solar
Instrumentation (S4I), which can be combined with a new and fast
polarimetry analysis. This new generation MSDP technique is well
adapted to large telescopes. Also presented are the goals of a
derived compact version of the instrument, the Solar Line Emission
Dopplerometer (SLED), dedicated to dynamic studies of coronal loops
observed in the forbidden iron lines, and prominences. It is
designed for observing total solar eclipses, and for deployment on
the Wroc{\l}aw and Lomnicky peak coronagraphs respectively for
prominence and coronal observations.
\end{abstract}
\end{opening}

\section{Introduction} \label{S-Introduction}

The study of active phenomena in the solar atmosphere, including
sunspots, filaments, prominences, flares or Coronal Mass Ejections
(CME) require high precision measurements of MHD parameters as
velocities and magnetic fields at various altitudes. In many cases,
a large field of view (FOV) and a high temporal resolution are
necessary to investigate dynamic events. Most instruments use either
tunable filters (Lyot or Fabry-P\'{e}rot) or thin slit
spectrographs, or both, such as the Solar Optical Telescope (SOT)
onboard Hinode \citep{Tsuneta2008}. Until 2016, the SOT/NFI filter
allowed the scanning of line profiles (with only a few points),
which in combination with polarimetry (FeI 6302 \AA, NaI D1 5896
\AA, MgI b1 6173 \AA) provided the evolution of the line of sight
(LOS) magnetic field. The SOT spectro-polarimeter (SP) records the
full Stokes parameters at high spectral resolution (30 m\AA) to
measure photospheric vector magnetic fields in FeI 6301/6302 \AA, at
low cadence because the slit must scan the solar surface. The
Helioseismic and Magnetic Imager (HMI, \cite{Schou2012}) onboard
Solar Dynamics Observatory (SDO) scans the FeI 6173 \AA~ line with
five points, in order to produce high cadence Dopplergrams and LOS
magnetograms of the photosphere. The Interface Region Imaging
Spectrograph (IRIS, \cite{Depontieu2014}) provides profiles of MgII
h and k chromospheric lines. Space-borne instruments are well suited
for line scans by tunable filters (no seeing effects); adaptive
optics is necessary for ground based telescopes, although residual
effects may affect the profiles.

The principle of the MSDP technique (developed by P. Mein since
1977) is the following: light from a 2D window is dispersed by a
first pass in the spectrograph; a beam splitter-shifter selects N
channels in the spectrum, and a second pass on the grating subtracts
the dispersion in order to form N spectra-images. They are not
monochromatic, because wavelength varies linearly in the x-direction
inside each channel, and by constant step between two consecutive
channels. It has the advantages of both filters (2D FOV and time
resolution) and spectroscopy (high spectral resolution). As the slit
is replaced by a rectangular window, spatial smoothing does not
exist. The MSDP provides data cubes (x, y, $\lambda$) where points
along the three coordinates are strictly simultaneous. It delivers a
large FOV at high spatial resolution and fast cadence observations;
this favours dynamic studies and magnetic field measurements (with
polarimetry) using broad chromospheric lines, and now weaker
photospheric lines with S4I. It is fully compatible with adaptive
optics and allows short exposure times. Doppler shifts and LOS
magnetic fields can be easily obtained at several altitudes using
the bisector method in line core and wings (as the core forms
higher), but more sophisticated inversion techniques have also been
applied to full line profiles.

The first MSDP instruments used prism-based beam-shifters (Meudon,
Pic du Midi, Wroc{\l}aw, VTT, THEMIS) and three subsequent
generations of instruments were produced with increasing
performances (spectral resolution from 300 m\AA~ down to 80 m\AA).
Our lastest generation beam-shifter (S4I) is now mirror-based in
order to improve the spectral resolution for photospheric lines (34
m\AA) and the number of channels for a better compromise between
spectral and FOV coverage.

Section 2 presents the historical background summarizing the
evolutions of previous MSDP spectrographs with prism beam-shifters.
In Section 3 we describe the S4I instrument using new generation
mirror slicers and associated polarimetric developments, together
with some preliminary results. Section 4 presents the derived
compact version designed for coronal forbidden lines (SLED). In
Section 5, we give, in the case of large telescopes, a short
comparison between new generation MSDP proposed for the European
Solar Telescope (EST) and instruments of the Daniel K. Inouye Solar
Telescope (DKIST). Section 6 is devoted to the discussion and
conclusions. A full MSDP bibliography is provided as on-line
supplement material.


\section{Historical Background of MSDP Spectrographs} \label{S-Historical-Context}

\subsection{Meudon Solar Tower: fast imaging spectroscopy of broad
lines (H$\alpha$)}

\begin{figure}
\centering
\includegraphics[width=0.5\textwidth,clip=]{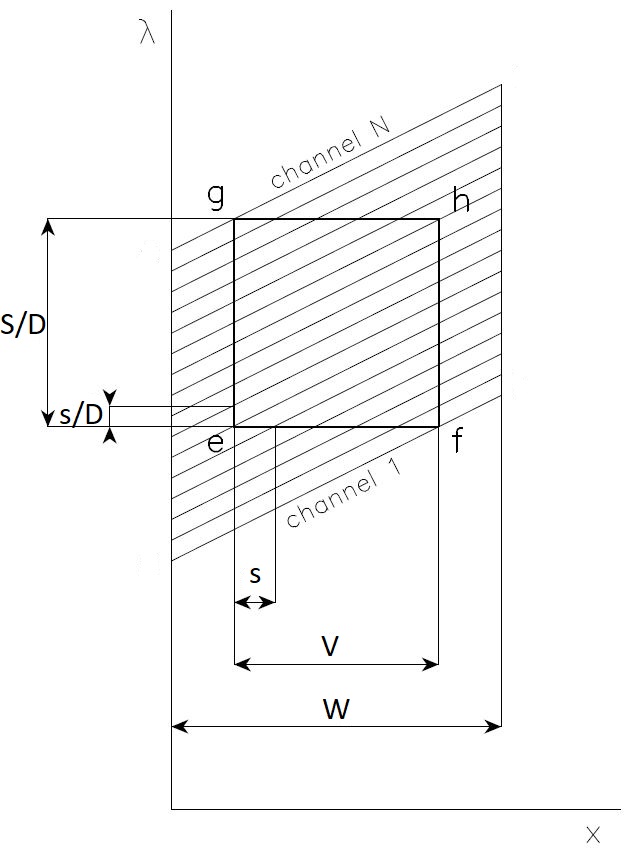}
 \caption[]{MSDP principle: N = number of channels; D = dispersion;
 s = multi-slit step; V = FOV in x-direction for a given spectral range S/D.
 } \label{principe}
\end{figure}

The first MSDP was installed in the Meudon Solar Tower
\citep{TOUR77}. At the telescope focus, a rectangular window (length
$L$ and width $W$ in the dispersion direction) replaces the usual
slit. After the first pass in the spectrograph (14~m focal length),
$N$ slits select $N$ wavelengths for each solar point, creating $N$
images after the subtractive second pass. The step $s$ between two
slits is equal to 2.5 mm, leading to the spectral resolution $s/D$
close to 0.30 \AA~ for the H$\alpha$ line, where $D$ (mm/\AA) is the
dispersion (Figure~\ref{principe}).

\begin{figure}
\centering
\includegraphics[width=1.0\textwidth,clip=]{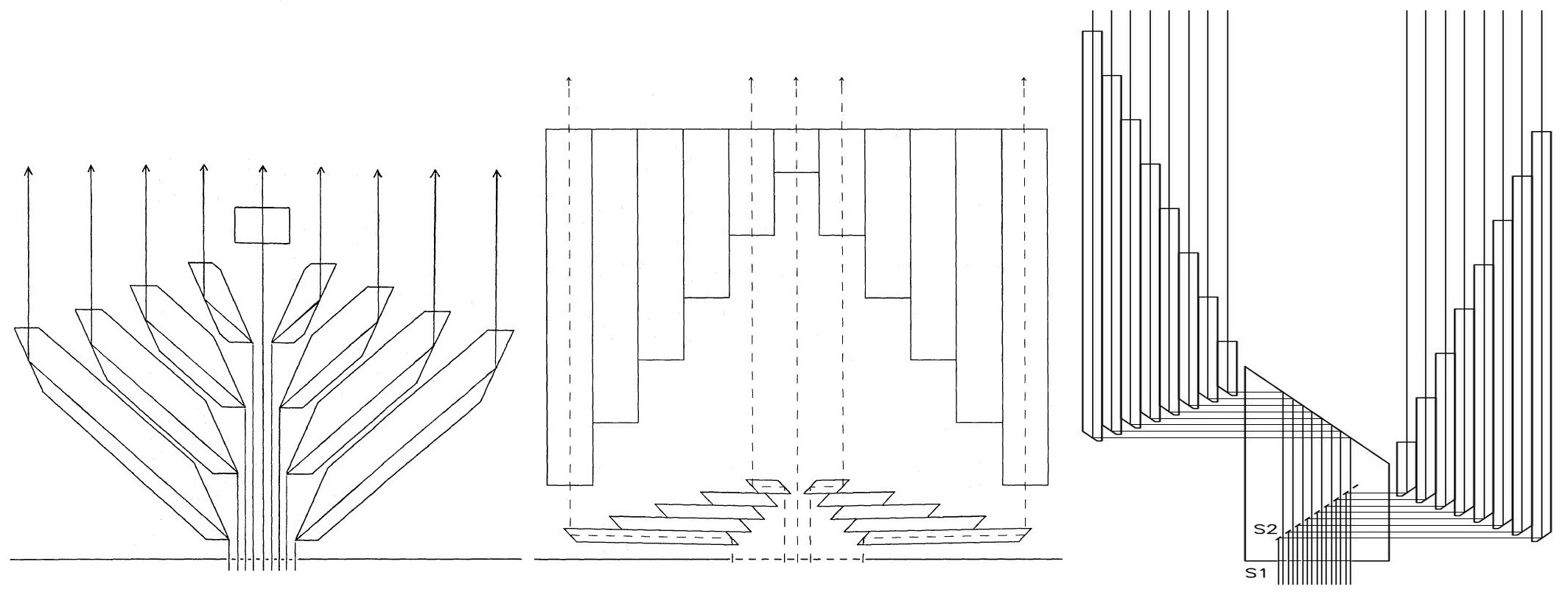}
 \caption[]{MSDP Beam-shifters. Left panel: 9 channels for broad lines (Meudon, Wroc{\l}aw and VTT);
 slit step 2.5 mm, slit width 1.5 mm. Middle panel: 11 channels for weaker lines (VTT);
 slit step 0.6 mm, slit width 0.3 mm. Right panel: beam-shifter with 16 channels used
 for high spectral resolution in the MSDP of the THEMIS telescope.
 The multiple slit S1 selects the wavelengths.
 Odd beams are reflected by the second multiple slit S2.
 The slit step is 0.4 mm and the slit width 0.2 mm.
 } \label{boites}
\end{figure}

Behind the slits, $N$ prisms shift the beams without focus
modification, in order to produce spatial distances between output
channels larger than the width of the entrance window  $W$ = 6 mm
(Figure~\ref{boites}a and Table 2). The number of channels,
initially 7, became $N$ = 9 in a 1979 version of the instrument. The
total width of the multiple slit is $s \times (N-1)$.

We now determine the maximum spectral range that can be recorded
across the full width $W$ of the window \citep{VTT91}. The $N-1$
intervals between slits can be divided into two parts $S = s \times
N_{1}$ and $V = s \times N_{2}$ (rectangle ``efgh'' in
Figure~\ref{principe}), corresponding respectively to the spectral
range $S/D$ (\AA) and to the width of the FOV (in the dispersion
direction), so that:

$s \times (N-1) = S + V = s \times (N_{1} + N_{2})$

\noindent To get a field of view equal to or larger than $W$  we
need:

$N_{2} = \frac{V}{s} > \frac{W}{s}$ = 6 mm / 2.5 mm = 2.4

\noindent With  $N_{2}$ = 3, we obtain $N_{1} = N - N_{2} - 1$ = 5
and $S = s \times N_{1}$ = 2.5 $\times$ 5 = 12.5 mm corresponding to
0.30 $\times$ 5 = 1.50 \AA~ (or $\pm$ 0.75 \AA~ from line centre)
for H$\alpha$ profiles.

When CCD detectors replaced photographic films, it was easy to
record about 120 pixels in the width of each channel. Taking into
account the Meudon seeing, it was reasonable to use only 2 pixels
per arcsec, and to place two lenses before the spectrograph entrance
to reduce the telescope focal length from 45  m to 22.5 m (Table 2)
leading to a 54$''$ field of view in the x-direction. At the same
time, a rotating system of prisms was used in front of the entrance
window to produce directly a scan of 5 successive images
\citep{Mein2008}. The total resulting FOV size is 270$''$ in the
dispersion direction (x). In the perpendicular direction (y), the
value of 440$''$ corresponds to a 50 mm length for the entrance
window.

As shown in Table 2, the scanning system can repeat scans of 270$''$
$\times$ 440$''$ FOV with a 30 s period. This is a very fast cadence
for spectroscopy and explains the high number of coordinated
campaigns which were organized between the Meudon MSDP and
radio-astronomy instruments or space solar probes. The immediate
availability of the Solar Tower for solar astronomers working in
Meudon was also clearly a major component of the success.

Fast imaging spectroscopy of H$\alpha$ is very well adapted to
dynamical chromospheric structures. Many publications were devoted
to oscillations of quiescent filaments \citep{Malherbe1981} and
chromospheric ejections connected with Type III radio bursts
(\cite{Mein1985}, \cite{Chiuderi1986}). Differential cloud models
were introduced to analyse the velocity fields of chromospheric
ejecta \citep{Mein1988}.

Coordinated observing campaigns, starting with the Nan\c{c}ay Radio
Heliograph, became increasingly frequent due to the space-borne
telescopes launched between 1980 and 2020. Some examples of topics
and publications resulting from these campaigns include:

\begin{itemize}

    \item Solar Maximum Mission/UVSP spectrometer: dynamics of filaments \citep{Malherbe1987}
    \item GOES, Yohkoh/SXT: surges and soft X-rays \citep{Schmieder1995}
    \item SOHO/SUMER and CDS UV spectrometers, SOHO/EIT (EUV imagery): dynamics of a slow ``disparition
    brusque''
          of a filament \citep{Schmieder2000}
    \item TRACE (high resolution EUV imagery), SOHO/EIT: Magnetic helicity in emerging flux
          and flare \citep{Chandra2009}
    \item Hinode/SOT, SOHO/SUMER spectrometers: 2D model of a prominence \citep{Berlicki2011}
    \item SOHO/SUMER: dynamics of prominence fine structures \citep{Gunar2012}
    \item SDO/AIA (EUV imagery): velocities in a coronal ``tornado''
    \citep{Schmieder2017}, long period oscillations in prominence
    \citep{Zapior2019}
    \item IRIS (UV spectrometer): dynamics of a quiescent prominence \citep{Ruan2018}
    \item IRIS, SDO/AIA: bidirectional reconnection outflows
    \citep{Ruan2019}

\end{itemize}

More references can be found in the full bibliography (supplement on
line material).

\subsection{Pic du Midi Jean R\"{o}sch Telescope: higher spatial and
spectral resolutions, polarimetry}

The high quality of the seeing at the Pic du Midi Observatory is
well known. A second MSDP  was installed in the  8 m spectrograph
that Z. Mouradian attached to the J. R\"{o}sch telescope in the
Turret Dome of Pic du Midi (\cite{Mein1980}, \cite{PIC89}). Two
beam-shifters were included, mainly for lines such as H$\alpha$ and
Na D1, with 11 channels, respective slit steps 1.2 mm and 0.6 mm and
slit widths 0.8 mm and 0.3 mm (Table 2).

The  Na D1 line allowed the observation of dynamical photospheric
fine structures. \cite{PIC03} reached the 0.3$''$ resolution for
photospheric granulation, close to the theoretical limit of the 50
cm telescope aperture. A similar spatial resolution was obtained for
circular polarization (\cite{Malherbe2004}, \cite{PIC06}). Strong
longitudinal magnetic fields could be associated with downward
velocities. Spectro-polarimetry with liquid crystals \citep{PIC07}
was used with high speed modulation between I+V and I-V signals
together with a fast detector, running in burst mode and allowing
image selection.

The H$\alpha$ line was also observed in arch filament systems
\citep{Alissandrakis1990}. Physical parameters of the plasma were
derived from a cloud model. Black and white mottles were also
investigated with non-local thermodynamic equilibrium (NLTE)  and
cloud models with non-uniform source functions (\cite{Heinzel1994},
 \cite{Tsiropoula1999}).

Coordinated observing campaigns were organized with:

\begin{itemize}

    \item Yohkoh/SXT: thermal flare \citep{Schmieder1998}
    \item SOHO/SUMER and CDS spectrometers, EIT imager: dynamics of a slow ``disparition brusque'' of a filament
    \citep{Schmieder2000}

\end{itemize}

\subsection{Tenerife German Vacuum Tower Telescope: two possible simultaneaous lines}

\begin{figure}
\centering
\includegraphics[width=0.6\textwidth,clip=]{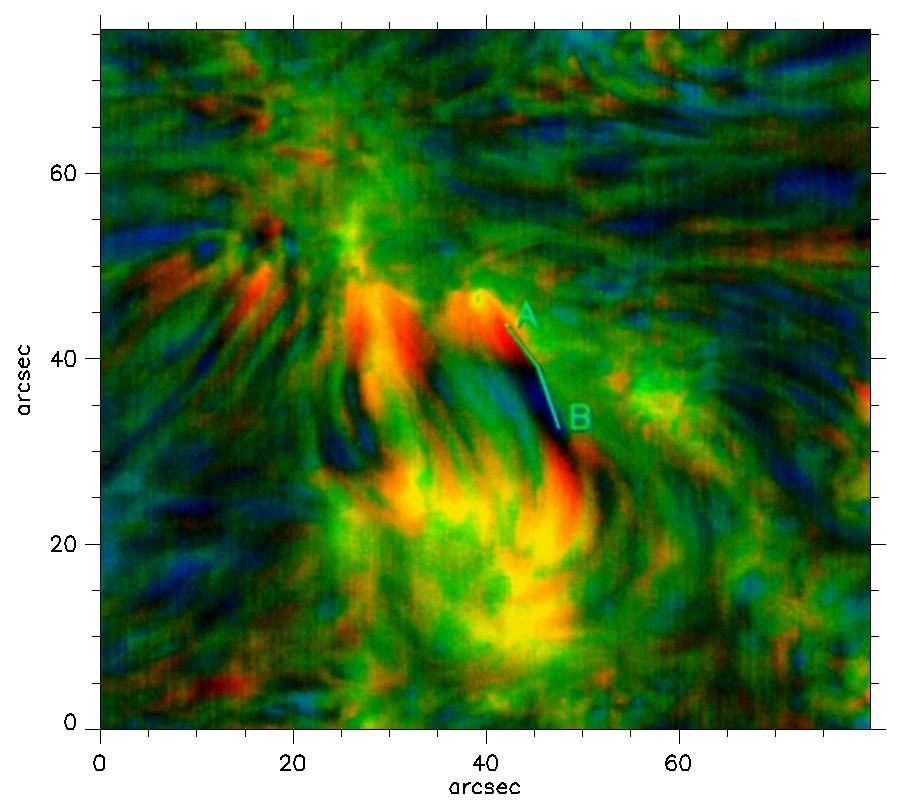}
 \caption[]{Arch Filament System (AFS) observed in H$\alpha$ on 5 October 1994
 with the MSDP/VTT. Intensities at line centre are shown in green. Doppler shifts (red/blue) are computed
 around the inflexion points of the profile ($\pm$ 0.45 \AA). In all arches such as (AB), the top is rising,
 while material is falling towards the footpoints which are anchored in bright photospheric faculae.} \label{afs}
\end{figure}

Thanks to J.P. Mehltretter from the Kiepenheuer Institute in
Freiburg, we were able to attach a new MSDP to the 15 m spectrograph
of the German Vacuum Tower Telescope (VTT) in Tenerife
\citep{VTT91}. Two prism beam-shifters with 9 and 11 channels (steps
2.5 mm and 0.6 mm) could be installed at the first focus of the
spectrograph, to record simultaneously two spectral lines, for
example H$\alpha$ and CaII 8542 \AA~ (Figure~\ref{boites}b). A cloud
model with variable source function was used to analyze the dynamics
of H$\alpha$ arch filament systems (Figure~\ref{afs}) reported by
\cite{Mein96}. Umbral flashes of sunspots and running penumbral
waves were observed simultaneously in both lines
\citep{Tziotziou2006}.

Coordinated campaigns were organized with:

\begin{itemize}

    \item SOHO/MDI: 3-D magnetic support of prominences \citep{Aulanier1998}
    \item TRACE, THEMIS/MSDP: Hydrogen density in emerging flux loops
                    \citep{Mein2001}
    \item TRACE, SVST (Swedish Vacuum Solar Telescope): Filament activation and magnetic reconnection \citep{Deng2002}
    \item SOHO/CDS and SUMER spectrometers: 3D structure of an EUV filament \citep{Schwartz2004}
    \item TRACE: Magnetic changes in formation of filaments \citep{Schmieder2004}
    \item SOHO/SUMER: Magnetic field orientation in a prominence \citep{Schmieder2007}

\end{itemize}

\subsection{Bia{\l}k\'{o}w Large Coronagraph: high time resolution and 3D models}

\begin{figure}
\centering
\includegraphics[width=1.0\textwidth,clip=]{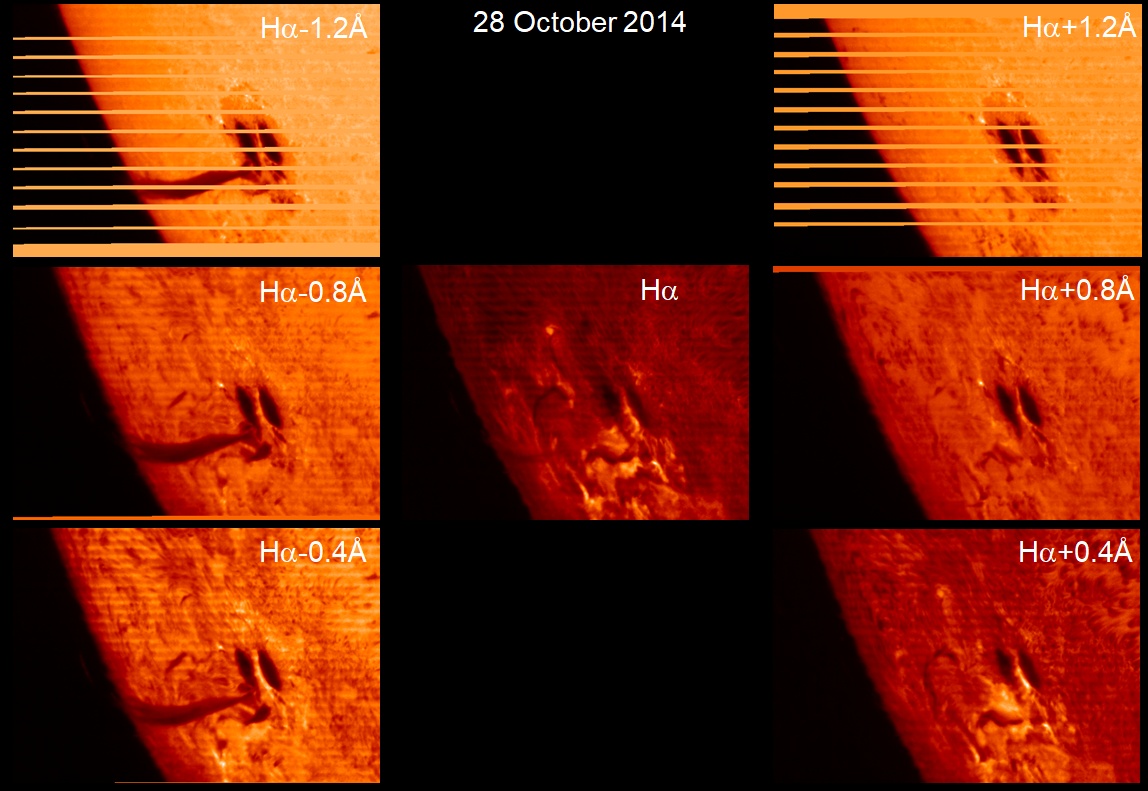}
 \caption[]{Surge observed with the MSDP spectrograph in the Bia{\l}k\'{o}w
 Observatory on 28 October 2014 in H$\alpha$ line center and
 wings ($\pm$ 0.4 \AA, $\pm$ 0.8 \AA, $\pm$ 1.2 \AA). The full FOV
 is made of 15 contiguous observations (exposure time 20 ms).} \label{surge}
\end{figure}

Within the framework of the French-Polish scientific cooperation,
the first MSDP, which was installed initially in the Meudon Solar
Tower \citep{TOUR77}, was relocated in 1993 to the Bia{\l}k\'{o}w
Observatory of the University of Wroc{\l}aw, Poland
\citep{Rompolt1994}. The MSDP was installed in the Coud\'{e} focus
of the Large Coronagraph \citep{Gnevyshev1967}, a classical
Lyot-type open-frame instrument having a 50.5 cm diameter main
objective aperture and nearly 14.5 m effective focal length,
replacing an original horizontal spectrograph of the Large
Coronagraph, but utilizing its massive substructure as well as the
collimator and camera mirrors. This instrument is mainly used for
prominence and chromospheric observations as the observatory
altitude is too low to observe the corona. If necessary, the
entrance window of the MSDP can be also fed by the 15 cm Horizontal
Telescope equipped with a 30 cm Jensch-type coelostat. The effective
time resolution of the collected data is up to 30 MSDP images per
second without spatial scanning; it takes about 23 s for a telescope
scan of 15 adjacent spectra-images covering a large FOV (300$''$
$\times$ 500$''$), as shown by Figure~\ref{surge}.

\begin{figure}
\centering
\includegraphics[width=1.0\textwidth,clip=]{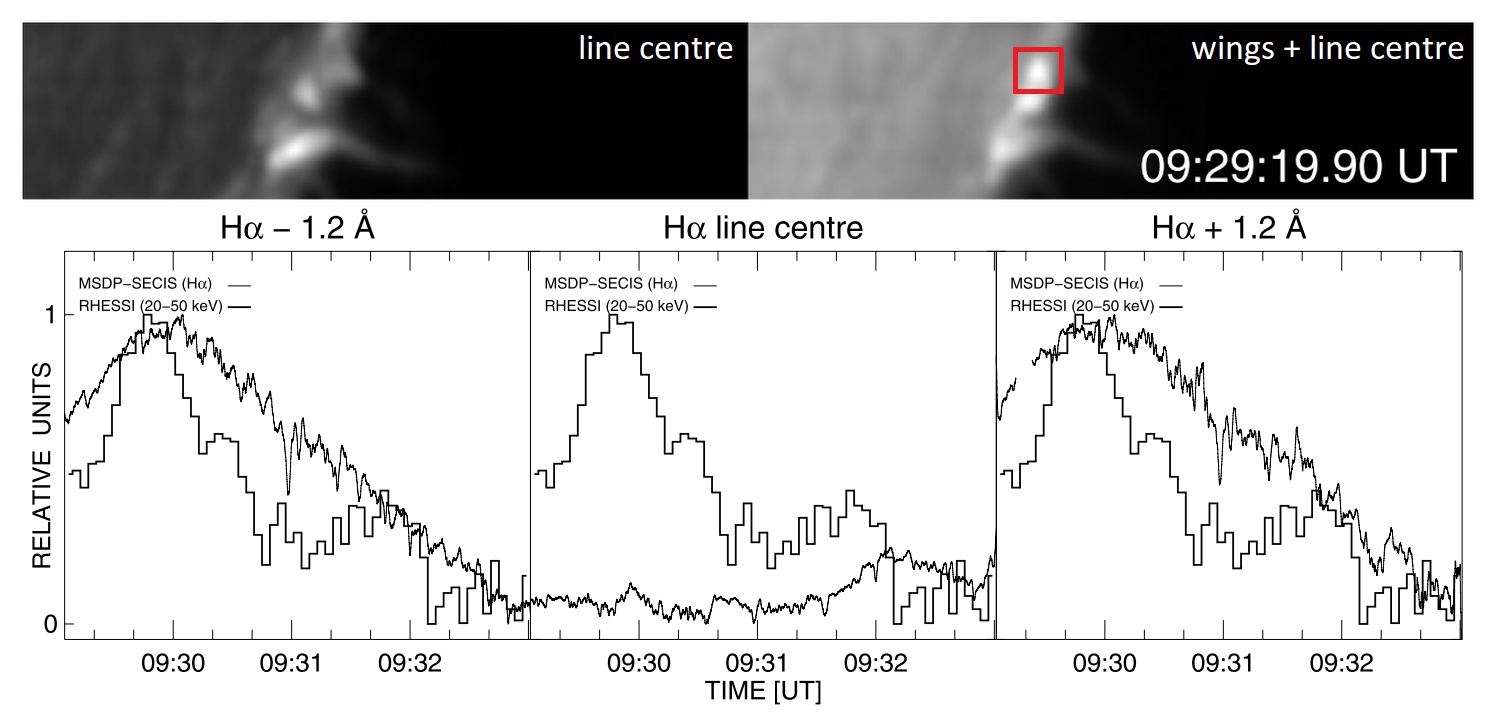}
 \caption[]{H$\alpha$ and X-ray light curves of a flare kernel (red box)
 observed on 23 April 2004 in the C4.4 GOES-class flare located
 at the west limb in active region NOAA 10 597.
 Observations in H$\alpha$ have a time resolution of 0.05 s. The H$\alpha$ emission
 is enhanced in the line wings only.
The 20-50 keV X-ray flux was observed by RHESSI. X-ray curves are
plotted logarithmically while H$\alpha$ data are displayed using a
linear scale.
 Top: H$\alpha$ center image (left) and sum of
 quasi-monochromatic MSDP images taken in the $\pm$ 1.0 \AA~ range (right).
 Bottom: X-ray and H$\alpha$ light curves (line core, blue and red wings).} \label{ker}
\end{figure}

Various solar phenomena were observed in Bia{\l}k\'{o}w. Fast
changes of H$\alpha$ emission in solar flares, correlated in time
with variations in hard X-ray fluxes, were investigated by
\cite{Radziszewski2007} using high time resolution (0.04-0.075 s)
MSDP observations of H$\alpha$ bright flaring kernels. These authors
reported very high correlations between variations of the X-ray flux
and H$\alpha$ emission from the selected bright flaring kernels
(Figure~\ref{ker}). The calculated precipitation depths of the
non-thermal electrons in the solar flares were validated using MSDP
observations of fast variations of the H$\alpha$ line profiles in
flaring kernels \citep{Falewicz2017}. Three-dimensional trajectories
of prominence knots were measured using data collected in the
H$\alpha$ line with a single telescope and without any assumption
concerning the shape of the trajectories or the mass motion
\citep{Zapior2010}. Applying restored 3D trajectories of the plasma
blobs, solar prominence magnetic fields were also estimated
\citep{Zapior2012}. The evolution and structure of numerous active
regions and solar flares were also studied (\cite{Rudawy2001},
\cite{Mandrini2002}, \cite{Rudawy2002}).

\subsection{Tenerife THEMIS telescope: two simultaneous lines, higher spectral
resolution and vector magnetic field measurements}

When the optical pattern of the THEMIS telescope was prepared by
Jean Rayrole, he decided to include a large predisperser in the 8 m
double spectrograph, in order to be able to select many lines
simultaneously for detailed studies of the photospheric magnetic
field. The Meudon group took this opportunity to use the
predisperser as a first pass on the grating, and the main
spectrograph as a second pass. Simultaneous observations with two
lines became much easier, and in addition scattered light effects
were strongly reduced. Several prism beam-shifters were built, with
slit steps of 1.2 mm (broader lines) and 0.4 mm (weaker lines).
Figure~\ref{boites}c shows the structure of the latter with 16
channels and 0.4 mm step. It provided a spectral resolution of
around 80 m\AA~ for polarimetry with the CaI 6103 \AA~ line
\citep{THEMIS02}. If (I, Q, U, V) is the Stokes vector, I$\pm$S
signals were observed simultaneously with S = Q, U, V alternately.
As the separation (provided by the calcite analyser) between I+S and
I-S was along the y-direction (orthogonal to the dispersion), it was
necessary to use grid masking, so that the FOV was covered by a 3 or
4 step scan along y-direction. The beam exchange technique was
available for deep polarimetry programmes.

\begin{figure}
\centering
\includegraphics[width=1.0\textwidth,clip=]{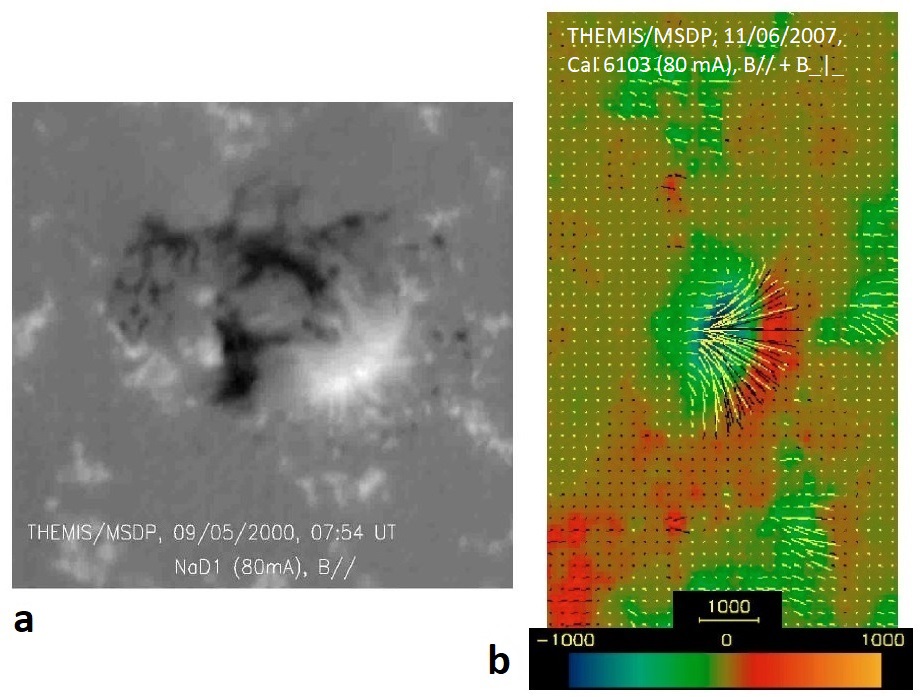}
 \caption[]{(a) LOS magnetic field from THEMIS/MSDP, FOV 138$''$ $\times$ 121$''$,
 9 May 2000.
 (b) Vector magnetic map derived from THEMIS/MSDP data and UNNOFIT inversion
 (BLOS in color; lines represent the module and direction of transverse fields);
 the FOV is 60$''$ $\times$ 105$''$. Observations taken on 6 June 2007.
 } \label{themis}
\end{figure}

Sunspot oscillations were investigated with the CaII 8542 \AA~ line
\citep{Tziotziou2002}. The vertical structure of spots, and the
magnetic models of flux tubes, were examined with the NaI D1 line
using two beam Stokes polarimetry with grid masking (LOS magnetic
field, Figure~\ref{themis}a, \cite{Eibe2002}, \cite{Mein2007}).
Spectro-polarimetric observations were performed for fine structures
of a filament in H$\alpha$, CaII 8542 \AA~ and NaI D2 5890 \AA~
\citep{Zong2003}. Fast vector magnetic maps (Figure~\ref{themis}b)
have been obtained with CaI 6103 \AA~ line \citep{THEMIS09}.

    Coordinated observations have been performed with:

\begin{itemize}

    \item SOHO/SUMER and CDS spectrometers: Extension of filaments in EUV and H$\alpha$  (\cite{Heinzel2001},
                   \cite{Schmieder2003})
    \item SOHO/MDI: Active region magnetic fields \citep{Berlicki2006}
    \item SOHO/MDI, TRACE: Ellerman bombs in CaII 8542 \AA, NaI D1 lines
                  (+ FeI 6302 \AA~ with Themis/MTR, \cite{Pariat2007})
    \item TRACE, VTT/MSDP: Hydrogen density in emerging flux loops \citep{Mein2001}

\end{itemize}


\section{The Spectral Sampling with Slicer for Solar Instrumentation (S4I)} \label{S-S4I}

\subsection{The new slicer}

\begin{figure}
\centering
\includegraphics[width=0.5\textwidth,clip=]{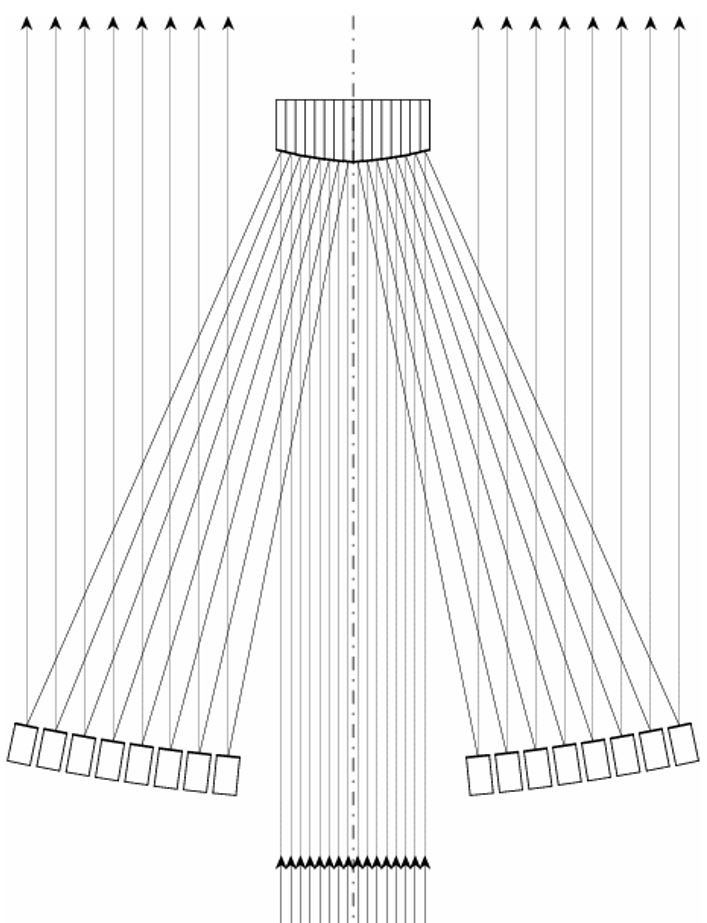}
 \caption[]{The mirror-based slicer of S4I provides more channels
 and improved spectral resolution (34 m\AA). Top: micro-mirrors
 (monoblock beam splitter). Bottom: shifting-mirrors (individually adjustable).
 } \label{slicer}
\end{figure}

\begin{figure}
\centering
\includegraphics[width=1.0\textwidth,clip=]{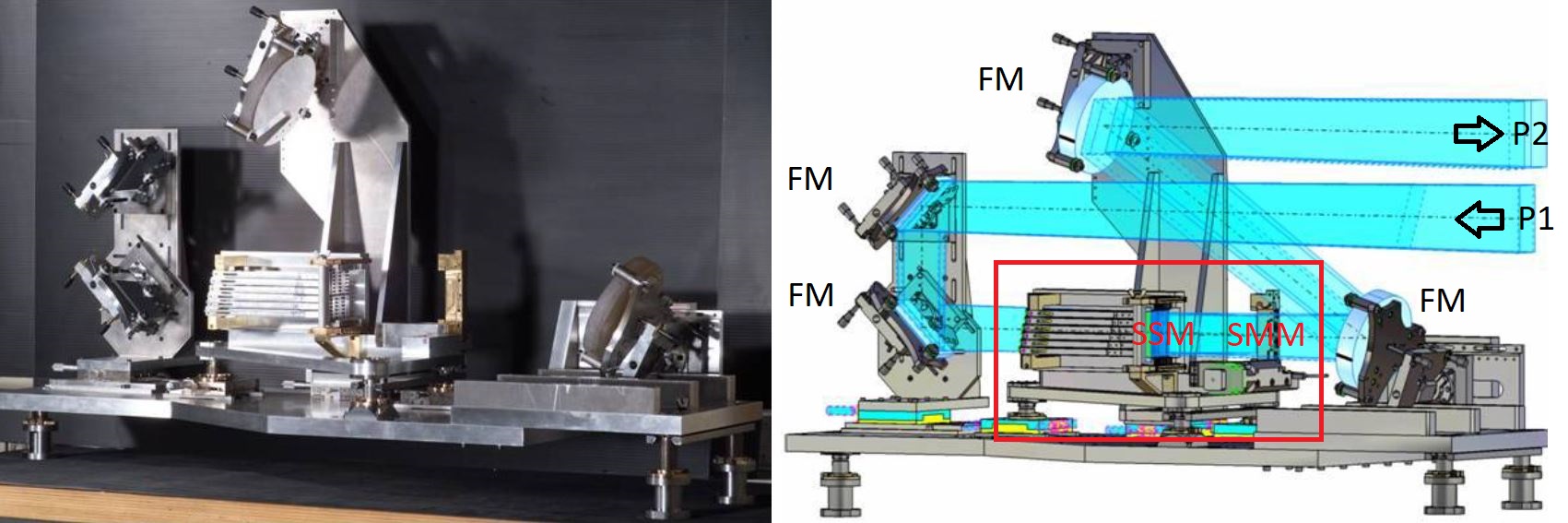}
 \caption[]{The S4I prototype of the Meudon 14 m spectrograph.
 P1 = light from first pass; P2 = light to second pass; FM = flat
 mirror; red box = slicer, composed of two parts:
 SMM = Slicer Micro-Mirrors (18 on the same substrate) and SSM = Slicer Shifting-Mirrors (18 in
 correspondence).
 } \label{S4Ileg}
\end{figure}

\begin{figure}
\centering
\includegraphics[width=1.0\textwidth,clip=]{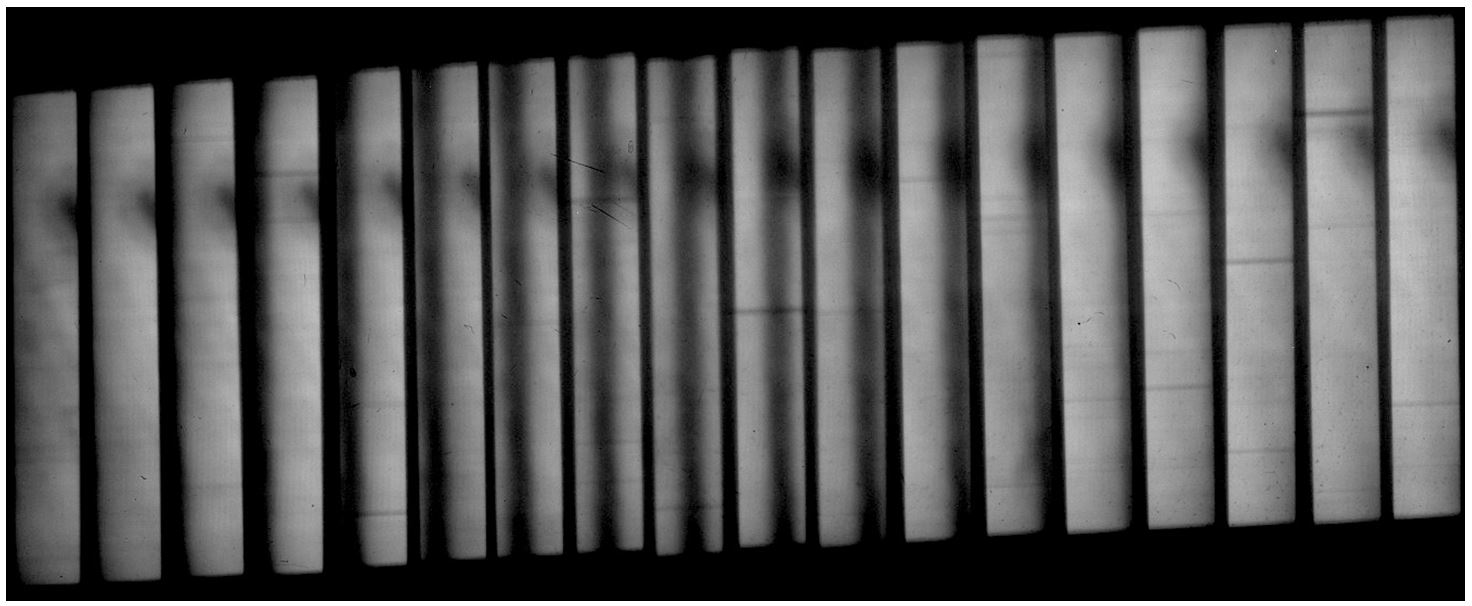}
 \caption[]{Observational test (under poor seeing conditions) with the S4I prototype of a sunspot
 in Mg\textsc{i} 5173 \AA~ line (34 m\AA~ spectral resolution). The
focal length of the telescope was reduced to 22.5 m (instead of 45
m), so that the FOV is 30$''$ $\times$ 250$''$ with 0.5$''$ spatial
sampling.
 } \label{obser}
\end{figure}

S4I is a prototype new slicer (Figure~\ref{slicer}) incorporated
into the 14 m spectrograph of the Meudon Solar Tower
(Figure~\ref{S4Ileg}). By contrast with previous devices using
prisms, S4I is based on micro-mirrors. This new concept
\citep{Sayede2014} provides an improvement in the spectral sampling
(34 m\AA) and the number of channels (18 at present, but easily
expandable to 30 or more) for a better compromise between spectral
($S/D$) and spatial coverage ($V$, see Figure~\ref{principe}). S4I
is designed for observations of broad chromospheric lines as well as
weaker photospheric lines (Table 2). Figure~\ref{obser} shows a
typical raw S4I image. The 18 channels are co-spatial, but are not
monochromatic, as wavelength varies linearly from the left to the
right border (the x-direction of Figure~\ref{principe}). The
wavelength shift between two adjacent channels is constant. It
corresponds to the spectral resolution (34 m\AA) and translates step
by step the spectral line in x-direction. The elementary FOV is
15$''$ $\times$ 200$''$. A scanning device (prismatic translator)
allows the observation of a 20 times larger FOV (300$''$ $\times$
200$''$) with 25 adjacent images obtained in sequence (12$''$ step).
Using a slow scanner and an old CCD camera (0.5 frame/s) available
at Meudon, the temporal resolution was only 160 s, but would be 10
times faster with modern sCMOS sensors and an optimized surface
translator. The camera of the prototype samples data with a hardware
resolution of 0.25$''$, but S4I is fully compatible with high
spatial resolution telescopes and adaptive optics.

The slicer is the core of the new MSDP and is composed of two parts:
\begin{itemize}
  \item N micro-mirrors acting as a multi-slit beam-splitter to select 2D channels
  in the focal plane of the spectrograph after the first pass on the
  grating. These N mirrors can be manufactured on a single substrate.
  \item N associated larger mirrors acting as a beam-shifter to form
  N spectra-images after subtraction of the dispersion by the second pass
  on the grating. These mirrors can be adjusted individually.
\end{itemize}

This new design provides a gain of a factor two in photon flux in
comparison to prism beam-shifters. S4I is also compatible with
larger aperture beams than previous image slicers (F/30 instead
F/60) and can now be integrated into compact and large FOV
spectrographs such as the SLED (see next section).

\subsection{New polarimetry with S4I}

\begin{figure}
\centering
\includegraphics[width=1.0\textwidth,clip=]{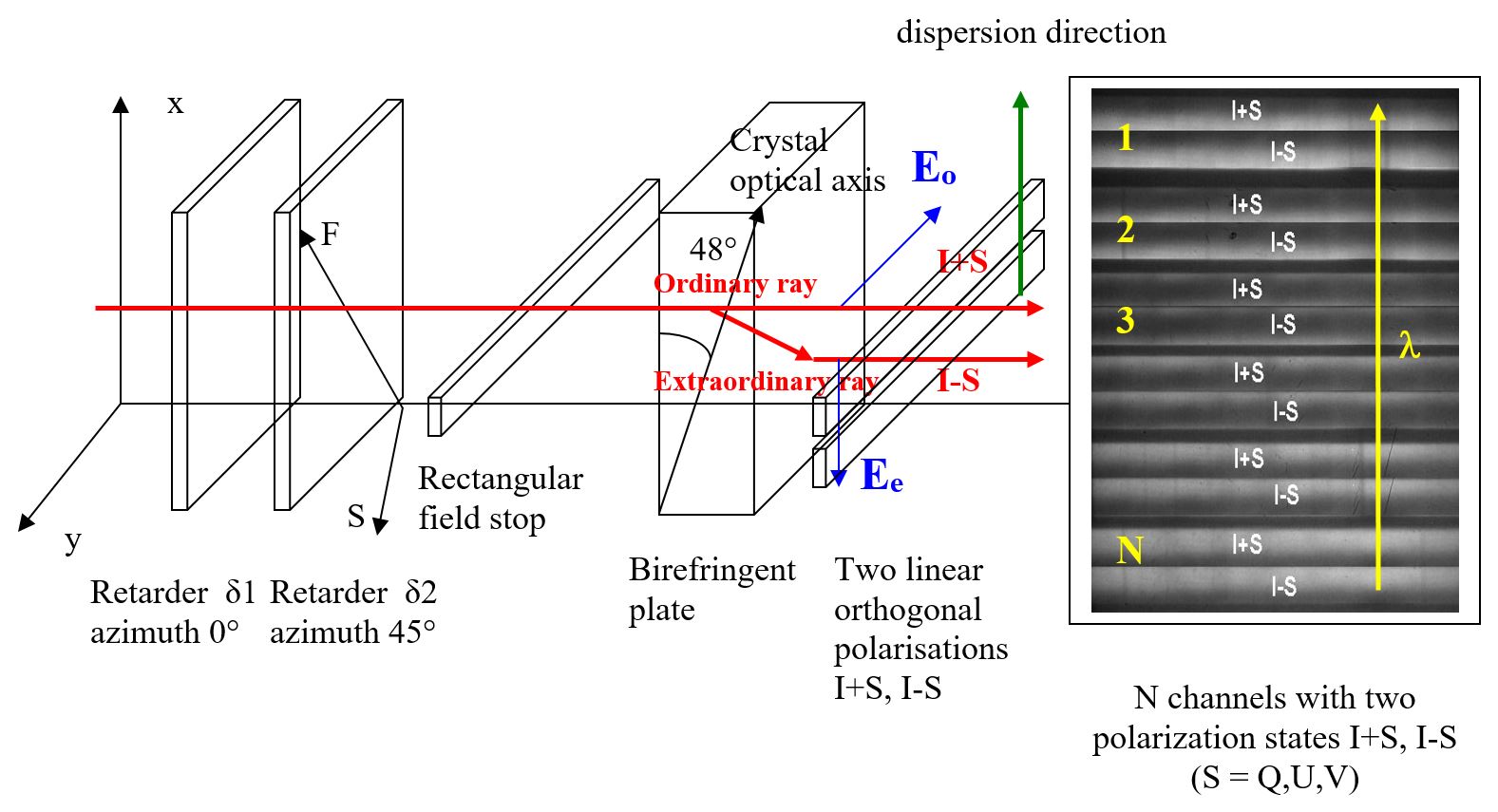}
 \caption[]{The S4I polarimetric analysis principle.} \label{polar}
\end{figure}

A new method for polarimetric analysis was developed and optimized
for S4I (Figure~\ref{polar}). The prototype polarimeter is located
in the image plane of the telescope. Two light beams are produced by
a calcite polarising beam-splitter (acting in the x-dispersion
direction), which provides two cospatial, simultaneous, and
orthogonally polarized sub-channels. These sub-channels are obtained
by reducing the width $W$ of the entrance field stop from 15$''$ to
7.5$''$. It is a fast method which has the advantage of suppressing
the y-scan (THEMIS polarimeter) and consequently distorsion in the
y-direction due to varying seeing. For Stokes V analysis, a single
retarder ($\delta_{1}$) is used, while for Stokes Q, U, V analysis,
two plates of retardance $\delta_{1}$ and $\delta_{2}$ are
necessary. The retarders are liquid crystals (manufactured by
Meadowlark Optics Inc.) which are precisely adjusted to the
wavelength of observations. The two signals:

\begin{itemize}

  \item $S_{1} = \frac{1}{2} ( I + Q \cos\delta_{2} + \sin\delta_{2} (U
\sin\delta_{1} - V \cos\delta_{1}) )$

  \item $S_{2} = \frac{1}{2} ( I - Q \cos\delta_{2} - \sin\delta_{2} (U
\sin\delta_{1} - V \cos\delta_{1}) )$

\end{itemize}

\noindent are observed simultaneously. For instance, $I \pm V$ are
obtained with $\delta_{1} = 0$, $\delta_{2} = \frac{\pi}{2}$; while
beam exchange is easy with variable retarders, $I \mp V$ are
provided by $\delta_{1} = 0$, $\delta_{2} = 3\frac{\pi}{2}$.
Combining beam exchange measurements may improve the polarimetric
accuracy, in the event of stable seeing. Stokes V was observed using
this method at Meudon, where U and Q are not reachable due to
instrumental polarization.

\begin{figure}
\centering
\includegraphics[width=1.0\textwidth,clip=]{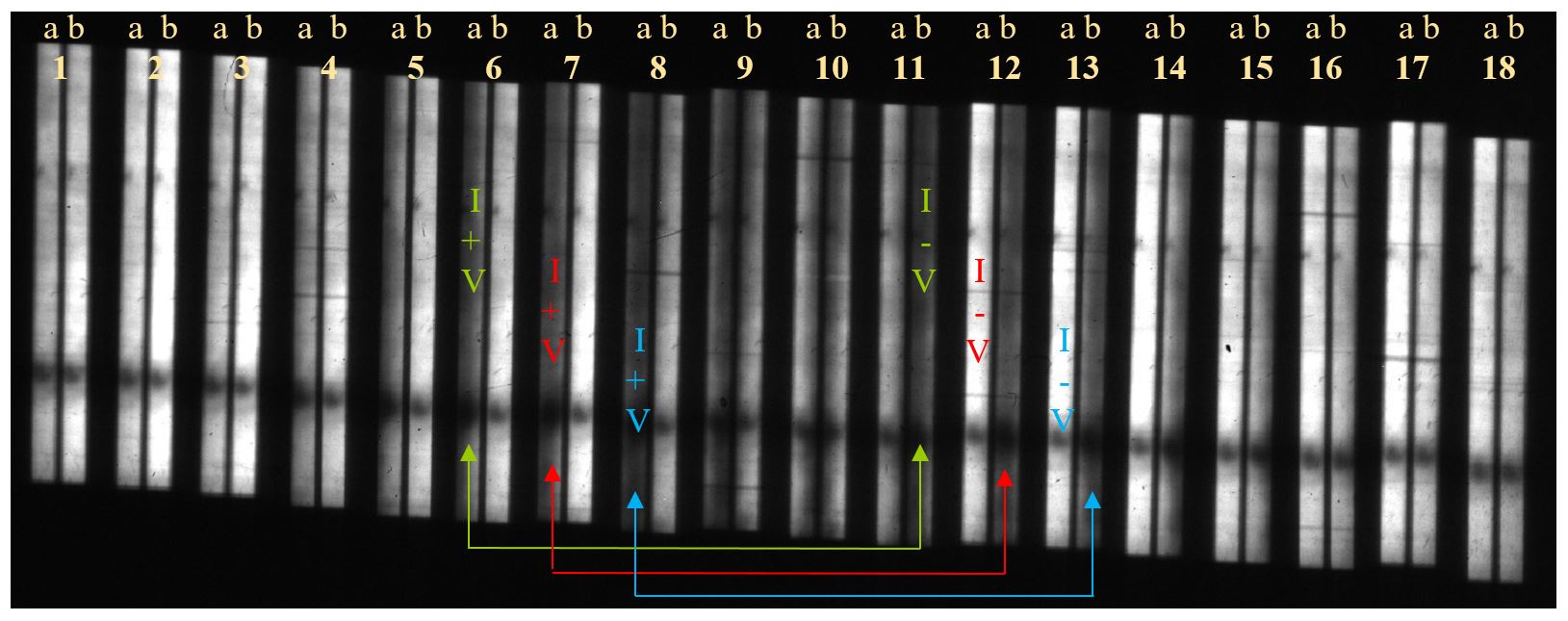}
 \caption[]{Polarimetric observation (I+V, I-V raw image) in Mg\textsc{i} 5173 \AA~ line. The
 elementary FOV is 12$''$ $\times$ 200$''$. Each channel (total of 18) is divided
 in two sub-channels (a,b) which
 are cospatial but wavelength translated (arrows).} \label{image}
\end{figure}

\begin{figure}
\centering
\includegraphics[width=1.0\textwidth,clip=]{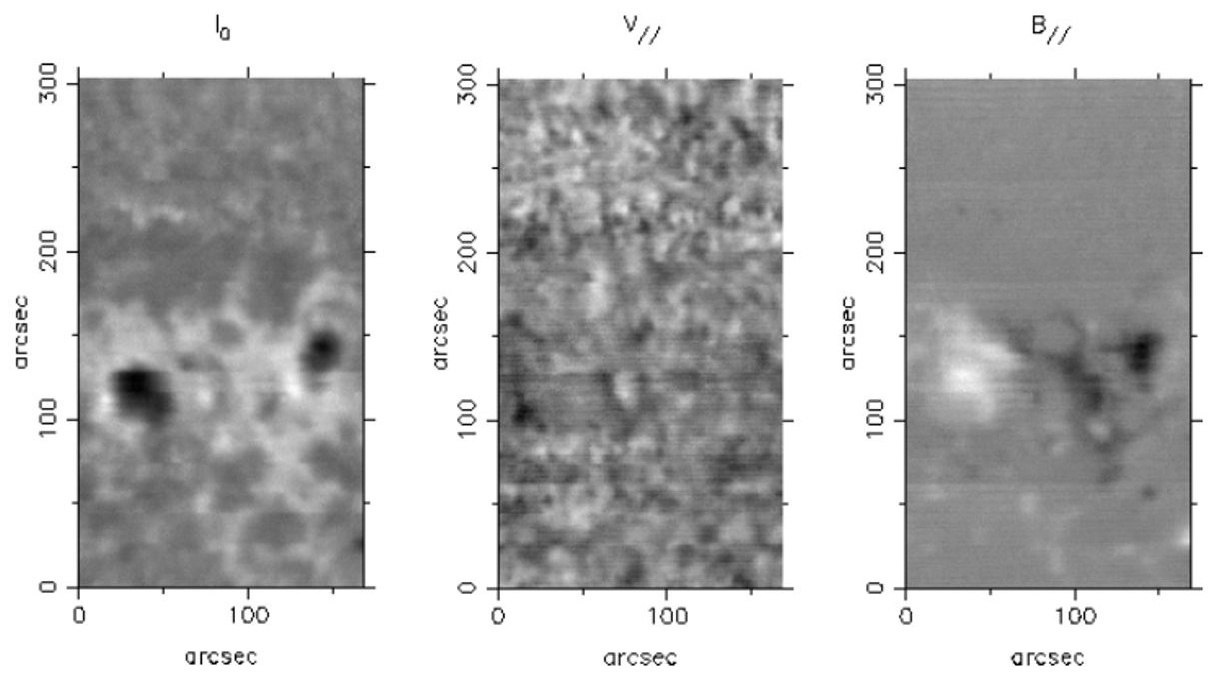}
 \caption[]{Intensity, Dopplergram and LOS magnetogram of a
 bipolar sunspot
 group obtained under poor seeing conditions; the full FOV of 170$''$ $\times$ 300$''$ is composed of 50
 elementary observations (6$''$ step).} \label{results}
\end{figure}

Polarimetric observations (Stokes V only) were tested at Meudon with
Mg\textsc{i} 5173 \AA~ line (Figure~\ref{image} and movie provided
as supplementary material). The raw image includes 18 channels, each
composed of two sub-channels (a, b); as wavelength varies along each
channel in the dispersion direction, the right sub-channel (I-V)
does not have the same wavelength as the left sub-channel (I+V), so
that (I+V) and (I-V) at the same wavelengths appear in different
channels (indicated by arrows). For instance, (I+V) of channel 6
corresponds to (I-V) of channel 11. Hence, the usable number of
wavelengths to cover the spectral line is reduced from 18 to 13. For
each pixel of the FOV, we derive, from line profile reconstruction,
intensities, Doppler shifts (the sum of (I+V) and (I-V) shifts with
respect to the quiet sun) and the LOS magnetic field (the difference
between (I+V) and (I-V) shifts). Preliminary results are displayed
in Figure~\ref{results}. A total of 50 successive images (taken at
6$''$ steps) are necessary to cover the final 170$''$ $\times$
300$''$ FOV; it takes 320 s with the old Meudon scanning device and
camera, but would be 10 times faster with modern hardware.

\section{The SLED} \label{S-SCD}

The Solar Line Emission Dopplerometer (SLED) is a new instrument
being developed by the Paris and Wroc{\l}aw observatories plus
Queen's University Belfast. It is a compact version (2.0 m focal
length, roughly five times shorter than predecessors) derived from
S4I for fast imaging spectroscopy of broad emission lines of the hot
corona, such as the iron forbidden lines FeXIV 5302 \AA~ (T $\geq$
1.6 $10^{6}$ K, ``green'' line) and FeX 6374 \AA~ (T $\leq$ 1.2
$10^{6}$ K, ``red'' line). The green line emission dominates at
temperatures above 1.3 $10^{6}$ K.

Monochromatic images in the green or red lines through interference
filters are recorded frequently by coronagraphs or during total
eclipse campaigns. The Wroc{\l}aw and UK groups have organized
several expeditions with the Solar Eclipse Coronal Imaging System
SECIS \citep{Phillips2000}, with a pair of fast CCD cameras working
at 70 frames/s. These lines have been also investigated by
spectroscopic means \citep{Singh1999} and low cadence
spectroheliograms have been obtained \citep{Singh2003}. However, to
our knowledge, the SLED will be the first instrument for fast
imaging spectroscopy dedicated to the green and red coronal lines.
Its scientific goal is to observe, through the Doppler effect, the
dynamics of hot loops, oscillations and mass motions in the low
corona at high temporal resolution. The SLED will search for
high-frequency wave-like variations of plasma velocities in active
region loops, which could be involved in coronal heating. In
particular, high-frequency waves (a few Hz) are suspected by theory
\citep{Porter1994} to dissipate readily at typical coronal
densities; the energy conveyed may (if the magnetic fields are
moderate) balance the energy losses of the corona. \cite{Rudawy2004}
and \cite{Rudawy2010} found, using SECIS images, slight evidence of
intensity oscillatory power in the range 0.05-1 Hz; the SLED will
produce fast cadence Dopplergrams to search for the signature of
high frequency (up to 10 Hz) wave-like motions in the corona.

The SLED is fully complementary with the Atmospheric Imaging
Assembly (AIA) onboard SDO, which provides continuously
monochromatic images from 80000 K to 10 million K in the EUV range
(94 - 335 \AA) at 45 s cadence.

\begin{figure}
\centering
\includegraphics[width=0.6\textwidth,clip=]{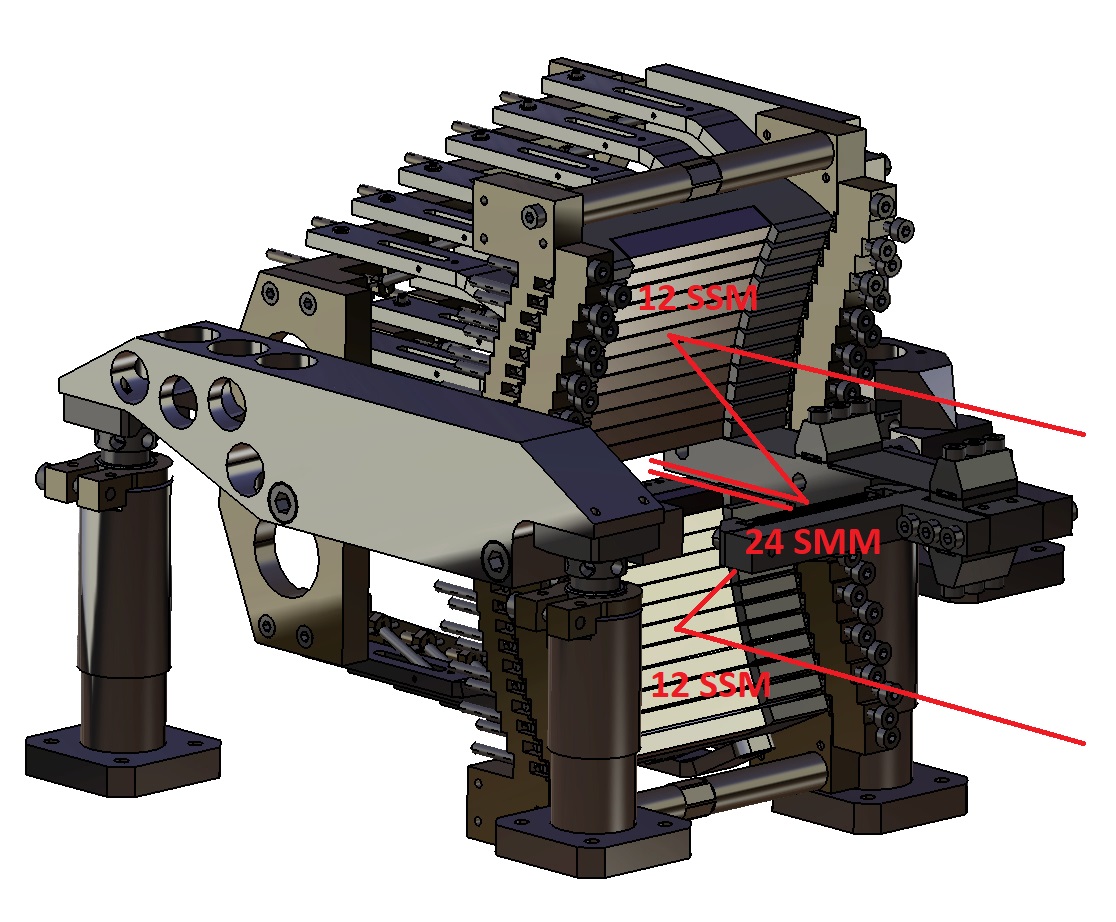}
 \caption[]{SLED slicer with 24 micro-mirrors (SMM) and 24 shifting-mirrors
 in two groups (SSM).} \label{slicerSCD}
\end{figure}

The slicer of the SLED (24 channels, Figure~\ref{slicerSCD} and
schematic view of Figure~\ref{slicer}) accepts beams at F/30 and is
now compatible with various existing telescopes. Hence, it is
envisaged to become a permanent instrument for the large
Bia{\l}k\'{o}w coronagraph (prominence observations) or another
high-altitude coronagraph, such as the Lomnicky Solar Observatory
(LSO) coronagraph (for hot lines), but it will also be used for
future total eclipse campaigns, as it is fully portable. The SLED
will be fed at F/30, at least by:

\begin{itemize}
  \item the 3.0 m focal length / 0.15 m aperture telescope (4.50 m equivalent
  length with Barlow 1.5 $\times$) for eclipse missions
  and heliostat, providing a 200$''$ $\times$ 1220$''$ FOV with 2.8$''$ pixel sampling
  \item the 3.0 m focal length / 0.2 m aperture coronagraph (6.0 m equivalent
  length with Barlow 2.0 $\times$) of the high
  altitude LSO, for coronal observations (hot plasma), 150$''$ $\times$
  1000$''$
  FOV with 2.1$''$ pixel sampling
  \item the 14.5 m focal length coronagraph (0.50 m aperture) at Bia{\l}k\'{o}w
  providing a 70$''$ $\times$ 220$''$ FOV (possibly larger with focus
  and entrance pupil reduction), for prominence observations (cold
  plasma)
\end{itemize}

\begin{figure}
\centering
\includegraphics[width=1.0\textwidth,clip=]{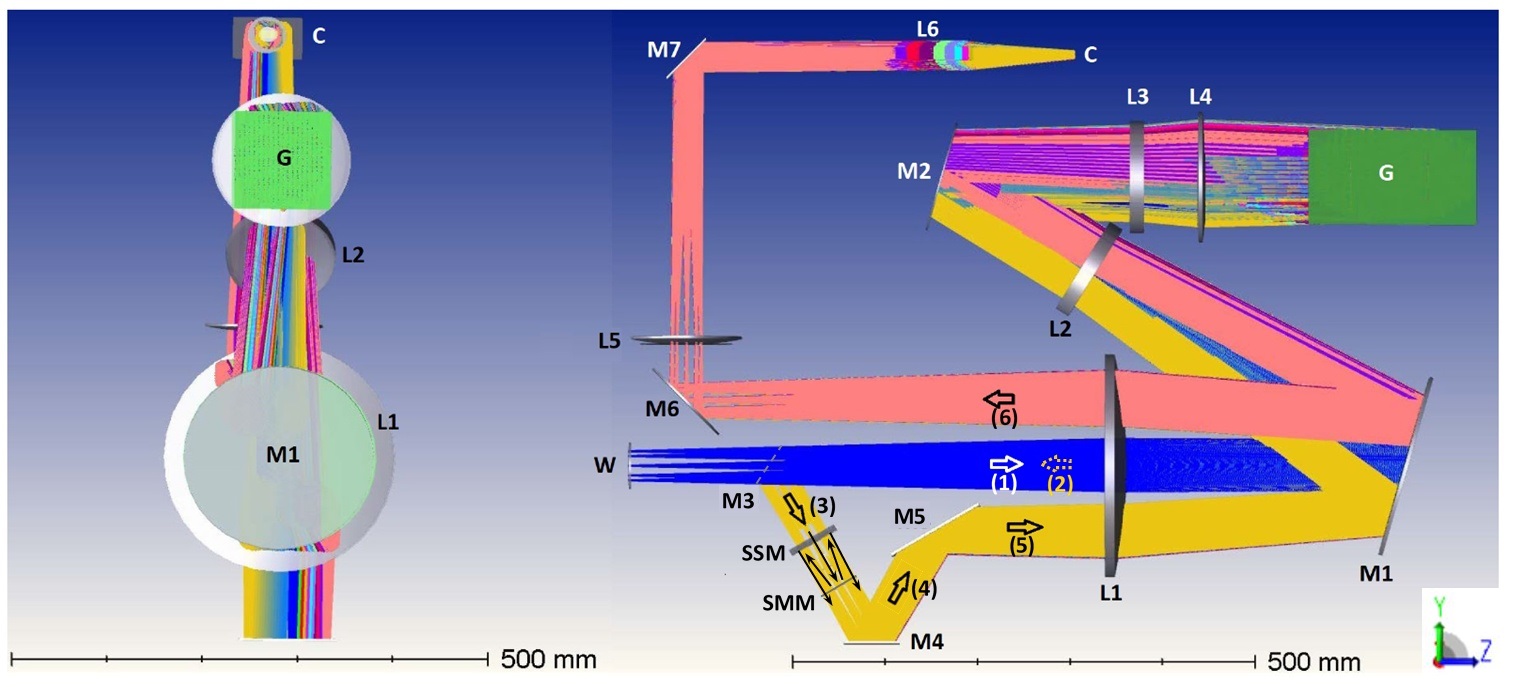}
 \caption[]{The SLED optical design; left is the front view, right the side view (YOZ plane).
  W = entrance window;
 L = lenses ; M = flat folding
 mirrors; G = grating; SMM/SSM = slicer micro-mirrors/shifting-mirrors;
 C = detector.} \label{design}
\end{figure}

The optical concept of SLED is shown in figure~\ref{design} and the
main characteristics are summarized in Table 2 for observations of
the hot corona. The spectral resolution (0.28-0.34 \AA) and spectral
coverage (24 channels, up to 6 \AA) is adapted to coronal line
widths (typically 0.9 \AA) and possibly high Doppler shifts (50 km
s$^{-1}$ $\approx$ 1 \AA). The large FOV (150$''$ $\times$ 1000$''$
with the LSO telescope, or more with the eclipse instrument) is
compatible with the size of coronal structures ($10^{5}$ km or
more). Its expected cadence is in the range 10-30 frames/s with a
fast sCMOS detector (depending on exposure time). The slicer is
built by Paris observatory; the spectrograph design, prepared in
Paris, will be assembled and tested by Wroc{\l}aw in the coming few
years.

The optical path of the SLED is enclosed in a 1.0 $\times$ 0.7
$\times$ 0.3 m$^{3}$ box and is indicated (for the green line at 530
nm) by the arrows of figure~\ref{design} (L = lenses or objectives,
M = flat mirrors):

\begin{itemize}
  \item W = rectangular entrance window (image plane, 4.4 mm $\times$ 29.0 mm)
  \item First pass (dispersion): W, L1, M1, L2, M2, L3, L4, G (grating in pupil
  plane), L4, L3, M2, L2, M1, L1, M3, SMM (slicer part 1)
  \item SMM/SSM = 24 channels slicer located in the spectrum (Figure~\ref{slicerSCD}).
  It is composed of
  two parts: 24 Slicer beam-splitting Micro-Mirrors (SMM) and 24 associated Slicer beam-Shifting Mirrors (SSM)
  \item Second pass (subtractive dispersion): SSM (slicer part 2), M4, M5, L1, M1, L2, M2,
  L3, L4, G (grating), L4, L3, M2, L2, M1, L1
  \item Transfer optics to the detector: M6, L5 (field lens, 600 mm focal length), M7, L6
  (100 mm focal length objective, 0.2 magnification, mounted on the camera)
  \item C = Andor Zyla camera (5.5 Mpixels sCMOS detector, 2560 $\times$ 2160, 6.5 micron
  square pixels, USB3, Peltier + air cooled), recording 24 channels
  spectra-images at 30 frames/s
\end{itemize}

\begin{figure}
\centering
\includegraphics[width=1.0\textwidth,clip=]{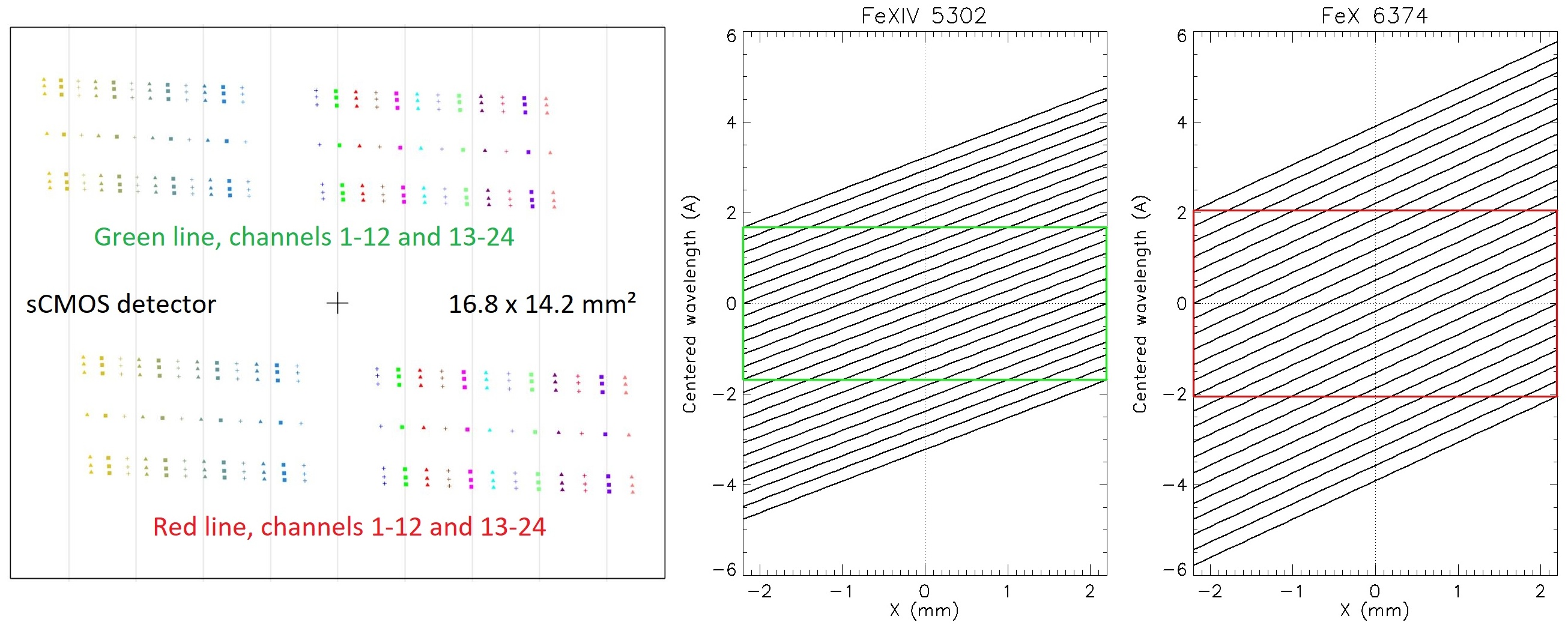}
 \caption[]{SLED detector patterns. Left: location of the 2 $\times$ 24 channels of the forbidden green
 and red lines on the detector (the axis of each channel is indicated by various symbols);
 right: spatial range (mm, 1$''$ = 0.0157 mm) and spectral range (\AA) covered by the 24 channels
 for FeXIV 5303 \AA~ (green box) and FeX 6374 \AA~ (red box).} \label{canaux}
\end{figure}

The optical combination of lenses (L1, L2, L3, L4) acts as
collimator and chamber objectives (2.0 m equivalent focal length)
with folding mirrors (M1, M2). The grating (G, 100 $\times$ 210
mm$^{2}$) is $62^{\circ}$ blazed and 79 grooves/mm ruled, and the
second pass on the grating subtracts the dispersion after selection
of $N$ = 24 channels by the slicer. Interference filters (not
represented) select the order (35 and 42 respectively for the red
and green coronal lines). L1, L2, L3 and L4 are the optical system
acting as a collimator in directions 1 (first pass) or 5 (second
pass), and as a chamber in direction 2 (first pass) or 6 (second
pass). The slicer (beam splitter SMM and shifter SSM) is shown by
Figure~\ref{slicer}, and the step of the SMM is $s$ = 0.4 mm
corresponding to $S/D$ = 0.34 \AA~ (R = 22000) and 0.28 \AA~ (R =
19000) spectral resolution, respectively for FeX 6374 \AA~ and FeXIV
5302 \AA~ lines ($D$ = dispersion, respectively 1.18 and 1.43
mm/\AA). In the case of an equivalent focal length of 6.0 m (LSO
coronagraph with Barlow 2.0 $\times$), the width of the FOV is
150$''$, corresponding to $V$ = 4.4 mm (width of entrance window),
with 2.1$''$ pixel sampling. We need to dedicate $V/s$ = 11 channels
to the FOV (Figure~\ref{principe}). As a consequence, 13 channels
remain available for line profiles. This corresponds to spectral
ranges $S/D$ = 4.1 \AA~ and 3.4 \AA, respectively, for the red and
green line (Figure~\ref{canaux}), allowing the determination of
large Doppler velocities. Both lines are observed simultaneously on
the detector (2 groups of 24 channels, Figure~\ref{canaux}). Overall
transmittance of the SLED is 0.25, and the expected number of
photons is proportional to the Coronal Emission Coefficient (CE);
for the green line, it is in the range 5-60, according to the phase
of the solar cycle. We estimate that errors on Doppler shift
measurements, for 0.1 s exposure time (10 frames/s), will be 0.9
km/s for CE = 10 and 0.4 km/s for CE = 50.

\begin{figure}
\centering
\includegraphics[width=1.0\textwidth,clip=]{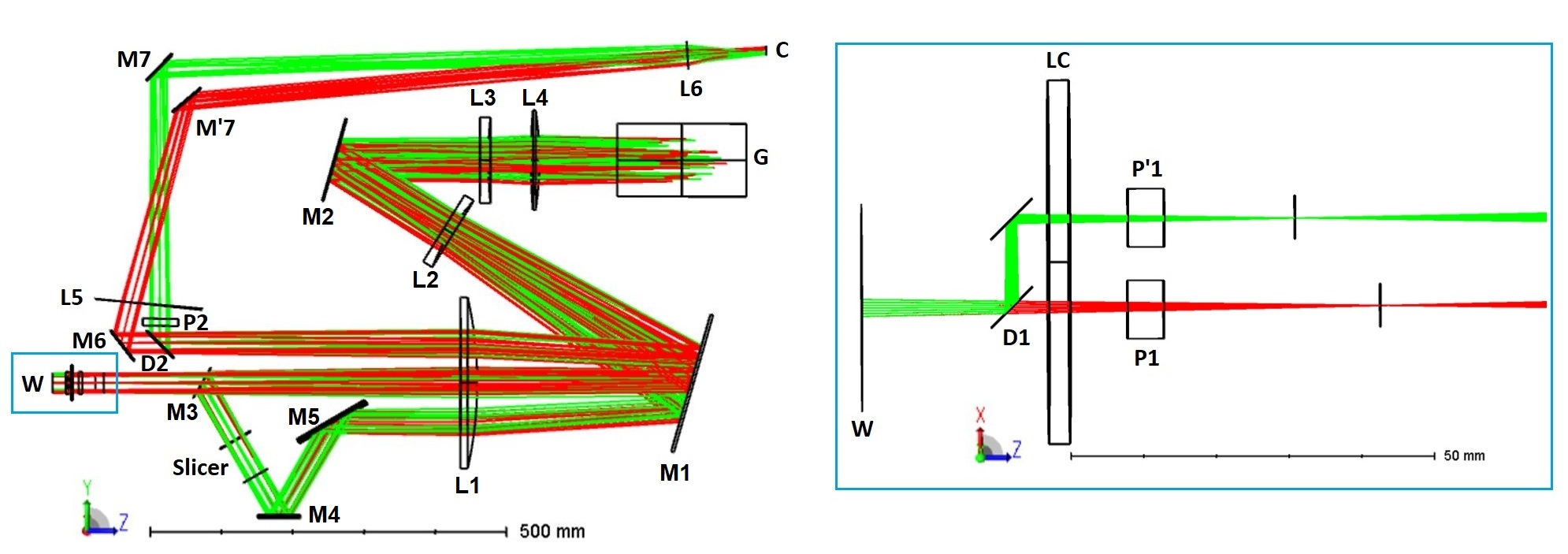}
 \caption[]{SLED optical path of FeXIV 5303 \AA~ (green line) and FeX 6374 \AA~ (red line).
 Left: in YOZ plane;
 right: detail of the beam splitting at the spectrograph entrance, corresponding
 to the blue box W at the left, in the XOZ plane.} \label{designgr}
\end{figure}

Figure~\ref{designgr} shows how the SLED is designed to observe
simultaneously two coronal lines. At the spectrograph entrance W,
the dichroic mirror D1 splits the green and red parts of the
spectrum. LC is a field lens. P1 and P'1 are compensation plates,
which can rotate around X axis to focus the spectrum on the slicer.
At the exit of the spectrograph, D2 is a second dichroic mirror to
combine properly on the same detector the FeX and FeXIV spectral
images. P2 is a plate for differential focusing on the camera.

Forbidden coronal lines can be observed at LSO. However, the SLED
can observe in H$\alpha$ (order 34) or He D3 5876 \AA~ (order 38)
lines to study at high cadence (30 frames/s) the dynamics of
prominences in the corona (respectively 0.35 \AA~ and 0.31 \AA~
spectral resolution). With the 14.5 m focal length of the
Bia{\l}k\'{o}w coronagraph, the width of the FOV will be 70$''$ but
could be increased to 140$''$ or more with focal reduction by a two
lenses Galilean system (reducing the image size without any
translation, as at Meudon Solar Tower). It would improve
considerably the capabilities of the MSDP already operating in
Bia{\l}k\'{o}w by suppressing the surface scan.

\section{New generation MSDP compared to IFU spectrometers and Fabry-P\'{e}rot filters for large telescopes} \label{S-Comparison}

We take advantage of the four decades of MSDP advances to discuss a
brief comparison with capabilities of new MSDP/S4I, Integral Field
Unit (IFU) spectrometers and Fabry-P\'{e}rot filters in the case of
four meters solar telescopes. We refer here to the MSDP project
proposed for EST \citep{Sayede2014} and compare it to the DKIST
instruments \citep{Rimmele2020} as the Diffraction Limited Near
Infrared SpectroPolarimeter (DL-NIRSP,
\url{https://nso.edu/telescopes/dkist/instruments/dl-nirsp/}) and
the Visible Tunable Filter (VTF,
\url{https://nso.edu/telescopes/dkist/instruments/vtf/}). Main
capabilities are listed in Table 1. Although all instruments are
able to perform polarimetry, we do not detail the corresponding
parameters. For MSDP, the spectrometer focal length is 8 meters with
aperture F/40. Widths of slicer mirrors are between 0.180 and 0.500
mm, and 56 channels are produced for each line. Filling factors are
commonly 90\%, but it must be noticed that the FOV and the spatial
sampling are quite dependent on the size of available detectors,
assumed to be 2 K $\times$ 4 K pixel format.

\begin{table}
\begin{tabular}{cccc}
  \hline
    & DL-NIRSP  & VTF & MSDP/S4I  \\
    & (IFU spectrometer)  & (Fabry P\'{e}rot filters) & (New generation)  \\
  \hline
  FOV ($''$) for & 2.4 $\times$ 1.8 (pixel 0.030) & & 120 $\times$ 8 (pixel 0.065)\\
  instantaneous & 6.2 $\times$ 4.6 (pixel 0.077) & & \\
  line profiles & 27.8 $\times$ 18.6 (pixel 0.464) & & \\
  cadence (ms) & 33 & & 33 \\
  \hline
  Scanned  FOV ($''$) & 120 $\times$ 120 (pixel 0.464) &  & $\geq$ 120 $\times$ 120 (pixel 0.065)\\
  cadence (ms) & 50 between positions & & 50 between positions \\
  \hline
  FOV ($''$) for scanned lines & & 60 $\times$ 60 (pixel 0.0146) & \\
  cadence (s) & & 3.5 for 50 wavelengths & \\
   \hline
  Spectral coverage (nm) & 3 lines in the bands & Sequential line scans & $\geq$ 3 line profiles\\
      & 500-900, 900-1350, 1350-1800 & 520-870 & UV to IR\\
  \hline
  Spectral resolution & 125000 at 900 nm & 100000 at 600 nm & 45000-150000\\
  \hline
\end{tabular}
\caption{Summary of capabilities of existing DL-NIRSP and VTF at
DKIST, and MSDP project for 4 m telescopes.}
\end{table}

To get high spatial resolution and fast imaging spectroscopy on
small solar targets, the IFU spectrometer with all simultaneous
wavelengths is very efficient. For larger FOV with high temporal
resolution, VTF observations are also excellent, although observed
wavelengths are not simultaneous so that differential seeing effects
must be considered. MSDP data cubes offer a good compromise between
FOV, cadence and spectral sampling optimized for each line. The
spatial resolution takes advantage of simultaneous wavelengths,
without smoothing due to any slit width. New generation MSDP
observes line profiles in the large FOV 120$''$ $\times$ 8$''$ with
small pixels of 0.065$''$. Larger FOV can be obtained by surface
scans (8$''$ steps), with line profiles always observed
simultaneously in each position.


\section{Discussion and conclusion} \label{S-Conclusion}

\begin{figure}
\centering
\includegraphics[width=1.0\textwidth,clip=]{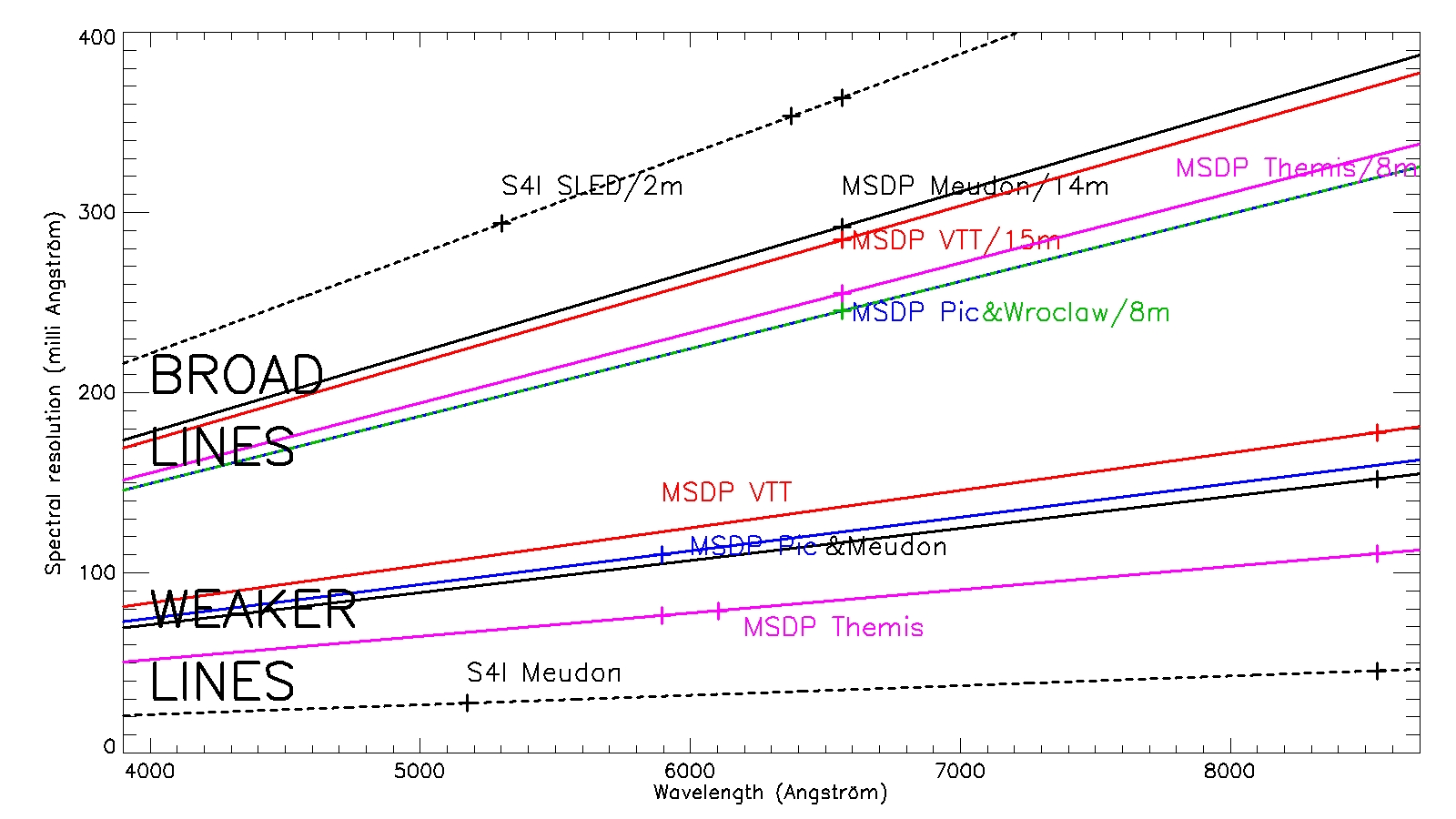}
 \caption[]{Spectral resolution (m\AA) as a function of wavelength (in the blaze)
 of the MSDP and S4I type instruments of Table 2. Crosses indicate
 some usual spectral lines. The spectrograph focal length is indicated.} \label{resol}
\end{figure}

Table 2 and Figure~\ref{resol} summarize the capabilities of the six
instruments which were produced from 1977 to the present. The core
of imaging spectroscopy is the beam splitter-shifter (or slicer);
the first generation (MSDP type in the table) was based on
multi-slits and prisms, providing convenient spectral resolution for
Fraunhofer lines. By contrast, the new generation of instruments
uses micro-mirrors (S4I type in the table), which are more compact,
improving the spectral coverage (24 channels or more) together with
the spectral resolution (up to 34 m\AA, convenient for a large
number of photospheric lines, Figure~\ref{resol}). With a native 2D
FOV, the technique provides, in comparison to classical
spectroscopy, a higher temporal resolution. This is particularly
well adapted to dynamic events of solar activity, or short duration
structures such as granulation. It is also compatible with adaptive
optics (high spatial resolution) and polarimetric analysis. While
THEMIS MSDP polarimetry used a grid, masking half of the FOV in the
y-direction orthogonal to the dispersion, we developed for S4I a new
and faster system avoiding masking. The calcite beam-splitter
separation (2.2 mm) operates now in the dispersion direction (x).
While three-step scans in y-direction were necessary at THEMIS to
compensate grid masking, this is no longer necessary with S4I.
Observations of a large FOV requires always x-direction scans, but
with big steps (typically 10$''$), either done by the telescope or a
motorized optical device near the image plane.

Some MSDP instruments have been decommissioned (Pic du Midi, VTT),
although THEMIS MSDP could be reactivated with adaptive optics.
Currently, instruments in operation are located in Wroc{\l}aw and
Meudon (both MSDP and S4I). SLED should commence operations in the
next few years. The Meudon data treatment software (Fortran) is
being revisited and will be available on-line, together with
documentation, at \url{http://bass2000.obspm.fr}. An IDL/Fortran
version exists at Wroc{\l}aw, available upon request (to P. Rudawy).
Meudon CCD observations will be freely available soon at the
BASS2000 solar database. Other or older data (Pic du Midi, VTT,
THEMIS) are archived on magnetic supports or photographic films.

\begin{table}

\begin{tabular}{cccccccc}
  \hline
  \textbf{Type} & MSDP & MSDP & MSDP & MSDP & MSDP & S4I & S4I  \\
  \hline
  \textbf{Instrument}     & Meudon   & Pic Midi & VTT & Wroc{\l}aw & THEMIS & Meudon & SLED \\
  \hline
   Slicer & Prisms & Prisms & Prisms & Prisms & Prisms & Mirrors & Mirrors \\
  Publication dates & 1977 & 1986 & 1991 & 1994 & 2000 & 2012 & $>$2021 \\
  (from/to) & 2019 & 2006 & 2007 &  -  & 2014 &  -   &   -   \\
  $F_{telescope}$ [m] & 22.5 (0) & 33-66 (1) & 45 & 14.5 & 60 & 22.5 (0) & 6.0 (2) \\
  CCD sampling [$''$] & 0.5 & 0.26-0.13 & 0.24 & 0.56 & 0.2 & 0.5 & 2.1\\
  $F_{spectro}$ [m] & 14.0 & 8.0 & 15.0 & 8.0 & 7.7 & 14.0 & 2.0 \\
  Storage (3) & F,CCD & F,CCD & F,CCD & CCD & CCD & CCD & sCMOS \\
  \hline
  \hline
  \textbf{Broad lines} & & & & & & & \\
  Example [$\lambda$ (\AA)] & 6563 & 6563 & 6563 & 6563 & 6563 & & 5303 \\
  Channels & 9 & 11 & 9 & 9 & 9 & & 24 \\
  Slicer step [mm] & 2.5 & 1.2 & 2.5 & 1.2 & 1.2 & & 0.4 \\
  Spectral step [\AA] & 0.30 & 0.26 & 0.28 & 0.40 & 0.27 & & 0.28 \\
  \hline
  \emph{Without polar} & & & & & & & \\
  x-scan step [$''$] & 54 & 25 & 15 (4) & 20 & 12 & &  \\
  x-value number & 5 & 6 & 6 (4) & 15 & 40 & &  \\
  xy FOV [$''$] (5) & 270$\times$440 & 150$\times$240 & 90$\times$120 &
  300$\times$500 & 480$\times$150 & & 150$\times$1000 \\
  Cadence [s] & 30 & 30 & 72 (4) & 25 (6) & 140 & & 0.04 (7) \\
  \hline
  \hline
  \textbf{Weaker lines} & & & & & & & \\
  Example [$\lambda$ (\AA)] & 5896 & 5896 & 8542 & & 6103 & 5173 & 6374 \\
  Channels & 9 & 11 & 11 & & 16 & 18 & 24 \\
  Slicer step [mm] & 1.0 & 0.6 & 0.6 & & 0.4 & 0.3 & 0.4 \\
  Spectral step [\AA] & 0.10 & 0.144 & 0.09 & & 0.08 & 0.034 & 0.34 \\
  \hline
  \emph{Without polar} & & & & & & & \\
  x-scan step [$''$] & & & 14 & & 5 & 12 & \\
  x-value number & & & 12 & & 26 & 25 &  \\
  xy FOV [$''$] (5) & & 14$\times$90 & 168$\times$180 &
  & 130$\times$160 & 300$\times$190 & 150$\times$1000 \\
  Cadence [s] & & 0.2 (7) & 60 & & 90 & 160 & 0.04 (7) \\
  \hline
  \emph{With polar} & & V (8) & & & VQU (9) & V (10) & \\
  Beam number & & 1 & & & 2 & 2 & \\
  Channels & & 11 & & & 16 & 2$\times$18 & \\
  y-scan step [$''$] & & & & & $\frac{1}{4}$ grid &  &  \\
  y-value number & & & & & 4 &  &  \\
  x-scan step [$''$] & & & & & 5 & 6 & \\
  x-value number & & & & & 26 & 50 & \\
  xy FOV [$''$] (5) & & 14$\times$90 & &
  & 130$\times$160 & 300$\times$190 & \\
  Cadence [s] & & 0.4 (7) & & & 700 & 320 & \\
  \hline
  \textbf{Simult. lines} & 1 & 1 & 1-2 & 1 & 1-2 & 1 & 2 \\
  \hline
\end{tabular}
\caption{MSDP (multi-slit beam splitter, prism beam shifter) and S4I
(mirrors beam splitter/shifter) based instruments. Remarks:
\textbf{(0)} with focal reducer ($\times$ 0.5); \textbf{(1)} with
focal magnifier ($\times$ 5 or 10); \textbf{(2)} with focal
magnifier ($\times$ 2.0);  \textbf{(3)} F = 70 mm photographic film
(4 cm $\times$ 6 cm images at Meudon or 6 cm $\times$ 9 cm images at
Pic du Midi and VTT), and later: frame transfer CCD, interline CCD
and finally sCMOS detectors; \textbf{(4)} scan step, x-value number
and cadence of the broad line depend on the elementary x-FOV of the
weaker line observed simultaneously; \textbf{(5)} (x, y) = (width,
length) of channels (x in the dispersion direction); \textbf{(6)}
much shorter (0.04 s or video rate) if no x-direction scan;
\textbf{(7)} fast burst mode with possible image selection and no
x-direction scan; \textbf{(8)} One beam (Liquid Crystal), I+V and
I-V sequential, 11 channels; \textbf{(9)} Two beams (Achromatic
Plates), I$\pm$V, I$\pm$Q, I$\pm$U sequential, beam exchange
possible, 4 step grid masking, 16 channels (3 measures for VQU, 6
measures with beam exchange); \textbf{(10)} two beams (Liquid
Crystal), I$\pm$V, 2 $\times$ 18 channels.}
\end{table}

Forty years of advances in imaging MSDP spectrometers have produced
more than 200 papers (full bibliography available as online
supplement material of the present paper). With a native 2D FOV,
previous MSDP (prism beam-shifters) and new S4I (mirror slicers)
technologies have many advantages, in comparison to single slit
spectroscopy, including both spatial (no slit smoothing) and
temporal resolution (a large FOV is covered by a small amount of
adjacent observations). Short exposure times (tens of ms),
eventually in association with adaptive optics (THEMIS), favour
image quality, which can be similar to that obtained through
monochromatic or tunable filters. Alternatively, fast cameras using
sCMOS detectors can be used to restore images by phase diversity
reconstruction from high speed bursts. The initial versions of MSDP
had spectral resolutions (typically 80-300 m\AA) adapted to
chromospheric lines. However, the new S4I design provides improved
spectral sampling (typically 30-40 m\AA) which allows observations
of weaker lines of the photosphere with long focal length
spectrographs (typically 10 m). A fast polarimetry technique was
also developed. The S4I prototype of Meudon is now miniaturized,
with new perspectives of much more compact and transportable
spectrographs (2 m) for either total eclipse campaigns, equipment of
high altitude coronagraphs (optical lines), or space missions (UV
lines). Hence, the SLED will be the first compact instrument
dedicated to coronal plasma dynamics, and will cover the hot green
and red forbidden lines, or cold prominence lines, when utilized
with coronagraphs in the next few years. The SLED design can work at
UV wavelengths as short as 1500 \AA, so that a space-borne version
for observing the CIV line of the transition region is under study.

\begin{acks}

We thank the referee for helpful comments and suggestions. We are
indebted to Ch. Coutard, R. Hellier, A. Miguel for the achievement
of MSDP instruments, and to Dr B. Schmieder and Dr N. Mein for
organizing and carrying out many observing campaigns. We thank also
D. Crussaire and R. Lecocguen for their technical assistance, and
the P\^{o}le Instrumental of the GEPI department (Observatoire de
Paris). We are grateful for financial support to the Institut
National des Sciences de l'Univers (INSU/CNRS), the Programme
National Soleil Terre (INSU/PNST), the University of Wroc{\l}aw, the
UK Science and Technology Facilities Council (STFC) and Queen's
University Belfast.
\end{acks}

\section*{Disclosure of Potential Conflicts of Interest}

The authors declare that they have no conflicts of interest.

\appendix

Electronic Supplemental Material.

\begin{description}

\item[i)] Movie (MPEG 4 format): a typical sequence of a solar scan with S4I
in polarimetric mode (25 successive images using a beam shifter are
recorded for a final 300$''$ $\times$ 150$''$ FOV; each image
presents 18 couples of sub-channels I+V, I-V. Each sub-channel
covers 300$''$ $\times$ 7.5$''$). First part of the movie:
Mg\textsc{i} 5173 \AA~ line; second part: Ca\textsc{ii} 8542 \AA~
line.

\item[ii)] MSDP full bibliography from 1977 to now (PDF format,
including URLs of the cited papers).

\end{description}

\bibliographystyle{spr-mp-sola}
\bibliography{PM_sola_V2}

\IfFileExists{\jobname.bbl}{} {\typeout{}
\typeout{****************************************************}
\typeout{****************************************************}
\typeout{** Please run "bibtex \jobname" to obtain} \typeout{**
the bibliography and then re-run LaTeX} \typeout{** twice to fix
the references !}
\typeout{****************************************************}
\typeout{****************************************************}
\typeout{}}

\end{article}

\end{document}